\documentclass[preprint]{aastex}

\usepackage{emulateapj5}
\usepackage{apjfonts}

\slugcomment{}

\shorttitle{EGRET upper limits from observations of galaxy clusters}
\shortauthors{Reimer et al.}

\begin{document}

\title{EGRET upper limits on the high-energy gamma-ray emission of
galaxy clusters}

\author{O. Reimer, M. Pohl}
\affil{Ruhr-Universit\"at Bochum, D-44780 Bochum, Germany}
\email{olr@tp4.rub.de, mkp@tp4.rub.de}

\author{P. Sreekumar}
\affil{ISRO Satellite Center, Bangalore, India}
\email{pskumar@isac.ernet.in}

\author{J.R. Mattox}
\affil{Department of Physics \& Astronomy, Francis Marion University,\\ Florence, SC 29501-0547, USA}
\email{JMattox@fmarion.edu}

\begin{abstract}

We report EGRET upper limits on the high-energy gamma-ray emission from clusters 
of galaxies. EGRET observations between 1991 and 2000 were analyzed at positions of 
58 individual clusters from a flux-limited sample of nearby X-ray bright galaxy clusters. 
Subsequently, a coadded image from individual galaxy clusters has been analyzed 
using an adequately adapted diffuse gamma-ray foreground model. 
The resulting upper 2 $\sigma$ limit for the average cluster is 
$\sim$ 6 $\times$ $10^{-9}$ cm$^{-2}$ s$^{-1}$ for E $>$ 100 MeV. 
Implications of the non--detection of prominent individual clusters and of the general 
inability to detect the X-ray brightest galaxy clusters as a class of gamma-ray 
emitters are discussed. We compare our results with model predictions on the high-energy
gamma-ray emission from galaxy clusters as well as with recent claims of an association
between unidentified or unresolved gamma-ray sources and Abell clusters of galaxies and 
find these contradictory.

\end{abstract}


\section{Introduction}

Clusters of galaxies are excellent representatives for the formation and the evolution of 
structure in the universe. They have been extensively studied at radio, optical and X-ray 
wavelengths. Within the last decade, radio, extreme UV and hard X-ray observations have 
revealed emission features that led to the prediction that galaxy clusters might be emitters 
of high-energy gamma-rays: 
\begin{itemize}
\item{the existence of diffuse radio halos \citep{gio93, gio99, gio00, kem01},} 
\item{the rather controversially discussed observations of EUV excess emission in 
galaxy clusters like A1795, A2199, and the Coma Cluster \citep{bow99}, 
Abell 2199 \citep{lie99}, A1367 and A1656 (Coma), A1795 and A2199 \citep{ara99}, 
Virgo \citep{ber00}, Virgo and A1795 \citep{bon01}, the Fornax Cluster \citep{bow01}, 
A2199 and A1795 \citep{ber02}, A1795, A2199, A4059, Coma and Virgo \citep{dur02},} 
\item{the observational hint of a distinct non--thermal emission component at hard X-ray 
wavelengths in the case of the Coma cluster \citep{fus99, rep99}, Abell 2199 \citep{kaa99}, 
Abell 2256 \citep{fus00}, and perhaps A754, A119 \citep{fus02}.}
\end{itemize} 
Various scenarios were suggested to connect and explain the links between these observations and,
consequently, to predict a high-energy emission component at gamma-ray wavelengths. Whereas the
diffuse radio emission is clearly synchrotron radiation by highly relativistic electrons, the 
EUV excess emission was first attributed to a second but cooler thermal component. Now a more
plausible explanation is Inverse Compton scattering of Cosmic Microwave Background radiation
by a non--thermal electron population \citep{ens98, bla99}. The hard X-ray excess can be produced 
by Inverse Compton scattering of the same electron distribution generating the non--thermal radio 
emission \citep{gio93}. To avoid the problem of the rather low magnetic field strength in such a
scenario, non--thermal bremsstrahlung has been proposed as an alternative emission process \citep{ens99}. 
As pointed out by \cite{pet01}, the non--thermal bremsstrahlung cannot be persistently produced on account 
of the low radiation efficiency of electrons in the 100 keV range. Hadronic particle populations were 
considered to produce gamma-rays via pp--interactions of high-energy cosmic rays with the 
intracluster medium (ICM) \citep{ber97}, or as the origin of a secondary population of 
relativistic electrons \cite{ato00}. Cluster merger systems might offer sufficient cosmic ray 
injection rates in conjunction with a mechanism for heating the ICM to the observed temperatures 
\citep{bla01, fuj02}.\\ 
Gamma-ray radiation from galaxy clusters is also expected as a result of large scale cosmological 
structure formation scenarios \citep{dar95, dar96, cola98, wax00, tot00, kaw02, min02}. 
However, apart from the general prediction of its existence, quantitative estimates range between 
'dominant part of the already observed extragalactic diffuse background by EGRET' to 
'magnitudes below the detection threshold of the current gamma-ray instrumentation' -
a range of predictions substantially more uncertain than that for the contribution of unresolved AGN to the 
extragalactic diffuse gamma-ray background (see i.e. \cite{mue00}, and references therein). 
The benefit of dealing with a class of astronomical objects already detected at gamma-ray wavelengths as 
e.g. AGN is not granted for the galaxy clusters: In contrast to the blazar population well-observed 
by EGRET, no galaxy cluster has been unambiguously identified at gamma-ray wavelengths to date.  
Nevertheless, for several individual clusters model predictions exist, which place their gamma-ray fluxes 
close to or even below the instrumental sensitivity threshold of the EGRET telescope at E $> 100$ MeV 
\citep{dar95, ens97, bla99}. Until now, galaxy clusters in gamma-rays have only been analyzed using early 
EGRET data and preliminary analysis techniques, resulting in non--detections of the Coma cluster \citep{sre96} 
and several Abell clusters \citep{gly94}. Therefore galaxy clusters have not been considered as likely 
counterparts of EGRET sources in the 3EG source catalog \citep{har99}.\\
Just recently, claims of an association between galaxy clusters from the Abell catalog and unidentified gamma-ray point 
sources from the 3EG catalog have been made by \cite{cola01} and \cite{kaw02}. 
Likewise, Abell clusters were proposed to be connected with unresolved gamma-ray excesses \citep{sch02}. 
All these detection claims have a statistical significance for association at the 3 $\sigma$ level in common.\\
Here, in order to provide an up--to--date and comprehensive view on the high-energy gamma-ray emission from
galaxy clusters, we have expanded a preliminary analysis by \cite{rei99} by considering all relevant EGRET observations 
between 1991 and 2000. Using the finalized EGRET data, which incorporate the latest instrumental efficiency 
normalizations, we analyzed individual, nearby X-ray bright galaxy clusters with the likelihood technique. 
Subsequently, the gamma-ray data from individual galaxy clusters have been coadded 
in cluster--centered coordinates \citep{rei01}. The coadded images were again analyzed using the 
likelihood technique, however in conjunction with an adequately adapted diffuse gamma-ray foreground model. 
We also re--examined the statistical associations between unidentified EGRET sources and Abell clusters as 
a population \citep{cola02}. For that purpose we measured the cluster autocorrelation and thus derived the correct
chance probabilities for the null hypothesis of no correlation between EGRET sources and Abell clusters. 
Finally, we compare our results, which benefit from the application of the likelihood analysis technique, with the 
result by \cite{sch02}.

\section{The flux-limited X-ray bright galaxy cluster sample}

For observationally probing the gamma-ray emission of galaxy clusters a sample of 
X-ray emitting clusters of galaxies has been chosen. This sample consists of the 
X-ray flux limited cluster catalogs from EINSTEIN \citep{edg90}, EXOSAT \citep{edg91}, 
and ROSAT surveys (XBACs: \cite{ebe96}, BCS north: \cite{ebe98}, BCS south: \cite{deg99}). 
Cluster selections based on X-ray catalogs currently provide the best way to 
obtain completeness without introducing biases (i.e. projection effects). 
Although appearing as extended sources with typical radii of several arcminutes in X-rays, 
the width of the point spread function of the EGRET instrument 
($\theta = 5.85^{\circ} (E_{\gamma}/100 {\rm MeV})^{-0.534}$, and $\theta$ the energy 
dependent radius for a 68\% flux enclosure) does not permit a similar handling of galaxy clusters as 
extended sources in gamma-rays. Thus, the attempt to analyze clusters of galaxies as point--like 
excesses at energies above 100 MeV is justified. Here, a total of 58 individual X-ray bright galaxy 
clusters within z$ < 0.14$ were chosen to represent a reasonable candidate sample for the subsequent 
analysis at high-energy gamma-ray wavelengths. Although further but similar cluster surveys are on 
the way or have been completed recently (in particular HIFLUGCS and REFLEX), this sample 
adequately represents the high-flux end of the log N--log S distribution of 
X-ray bright galaxy clusters. Almost all clusters that are extensively discussed 
in the literature for evidence of non--thermal X-ray emission, EUV-excess features 
and/or characteristic diffuse radio halos are included in this sample. 
The number of galaxy clusters has been restricted primarily to achieve a manageable 
amount of analysis work in the gamma-rays, but also because a simple measure 
like cluster mass M over distance squared D$^2$, as explained in the discussion, should 
already be major constraint for the detectability of galaxy clusters in gamma-rays. 
Thus only nearby clusters (z$ < 0.14$) were considered. This choice reflects the 
expectation that the nearest, most massive galaxy clusters are most likely the ones 
to be detected as individual sources of gamma-ray emission, whatever their flux.
Figure 1 shows the spatial arrangement of the galaxy cluster sample ($\times$)  
and cataloged high-energy gamma-ray sources \citep{har99}, in galactic coordinates.\\

EDITOR place figure 1 here\\

\section{Analysis of galaxy clusters at high-energy gamma-rays}
\subsection{The study of individual galaxy clusters}

Until recently, no positional coincidences between an individual galaxy cluster and gamma-ray point sources 
in existing EGRET source catalogues have been reported. For the Coma cluster the result of an EGRET analysis 
has been published, based on observations from CGRO cycle 1 and 2 \citep{sre96}. In the analysis described here, 
EGRET data of individual viewing periods from CGRO observation cycles 1 to 9 were used for the analysis of 
58 individual clusters. The latest and presumably final improvements in the efficiency corrections 
for the instrumental response of the EGRET spark chamber telescope have been fully implemented. 
Each galaxy cluster has been individually analyzed by means of standard EGRET data reduction techniques 
(likelihood source finding algorithm and subsequent flux determination at the position of the center of 
the X-ray emission). This analysis goes beyond the preliminary study presented by \cite{rei99}, in which four 
years of EGRET observations were analyzed in strict congruence with the four years of EGRET observations 
used for the 3EG catalog of gamma-ray point sources. Coadded images of individual viewing periods,
where a cluster has been observed at less the 30\degr\ off the pointing axis of the EGRET instrument (standard 
field-of-view observations) or less then 19\degr\ (narrow field-of-view observations), have been searched for 
gamma-ray excesses after modeling cataloged (and therefore well-known) identified gamma-ray point 
sources by using the maximum-likelihood technique as described in \cite{mat96}. Gamma-ray source fluxes have been 
determined at the coordinates of the cluster center position known from X-ray observations. Applying the same 
detection criteria as used and described in the EGRET source catalogs, none of the 58 galaxy clusters are 
detected in the EGRET data. Special care has been exercised when already cataloged gamma-ray point sources 
are near the position of an Abell cluster, see footnote of Table 1. Three of these sources are 
unambiguously identified blazar-class Active Galactic Nuclei: 1633+382 = 3EG J1635+3818 near A2199, 
3C279 = 3EG J1255-0549 near A1651, and 1604+159 = 3EG J1605+1553 near A2147. In these cases the particular AGN
has been modelled at it's known radio position and simultaneously taken into account in the deteremination of the 
gamma-ray flux at the position of the Abell cluster in question. Only one of the remaining three 
catalog sources shows considerable overlap at the position of an analyzed cluster (A85 with the unidentified 
source 3EG J0038-0949). Keeping in mind the width of the point spread function of the EGRET telescope, the 
total number of unidentified gamma-ray sources, and the size of our galaxy cluster sample, this occurrence 
is perfectly in agreement with chance coincidence: The probability of at least one coincidence between one 
of the 170 unidentified EGRET sources and one of the 58 considered galaxy clusters at a distance of 0.81\degr\, 
as in case of 3EG J0038-0949 and A85, is 48.1 \%. Here, a gamma-ray flux has been obtained at the cluster 
position without explicitly modeling the known, but unidentified gamma-ray source.\\ 
Exemplary for the detailed results given in Tab.1, the intensity and test statistics map of three prominent 
galaxy clusters (Perseus, Virgo, and Coma) are shown in Fig.2.\\

EDITOR place figure 2 here\\

The most significant gamma-ray excess for any individual cluster has been found for A3532, corresponding to 
a detection significance of 1.6 $\sigma$. This is well below the detection threshold for being seriously considered 
as a source by standards of the EGRET data analysis, which is $> 4 \sigma$ at high-galactic latitudes. 
Furthermore the achieved detection significance is less than expected from statistical fluctuations at 58 trials 
alone. Therefore, for each galaxy cluster a 2 $\sigma$ upper limit at the position of the cluster center has been 
determined and is given in Tab.1\\

EDITOR place table 1 here\\

\subsection{The study of the cluster population}

Having established that individual clusters are not found in the EGRET data, a further analysis has been 
performed to study whether or not these galaxy clusters radiate in gamma-rays as a population. For this purpose, 
the counts, exposure, and intensity maps of the individual galaxy clusters have been used. Each individual map has been 
transformed into a cluster--centered coordinate system under conservation of the original pixelation. Subsequently, 
the individual images have been coadded. Three sets of images have been produced in order to assure that the center 
region of the stacked image is not dominated by already identified point sources or the galactic plane:\\ 
(1) a superposition of all 58 galaxy clusters in the sample,\\
(2) a superposition of 54 galaxy clusters excluding those with unambiguously identified and dominant EGRET sources 
in the central map bins (A2199: 1633+382, A1650, A1651, and A1689: 3C279), and\\ 
(3) a superposition of 50 galaxy clusters excluding those with unambiguously identified and dominant EGRET sources (see above) 
or the Galactic Plane in the map center (the Oph and Cyg A cluster, 3C129, and 3A 0745-191).\\
  
Each of the three sets have been analyzed, however, here we only report from the least contaminated set (3).
Due to the wealth of accumulated instrumental exposure the results do not differ dramatically between these selections. 
The interpretation, however, simplifies considerably since we are not urged to discuss identified and dominant point sources 
or foreground emission features at or close to the image center anymore. The total exposure in the center region,
averaged over the four central 0.5\degr\ $\times$ 0.5\degr\ map bins, is $3.4\times 10^{10}$ cm$^2$ s (E $> 100$ MeV)
for the 50 cluster selection, the lowest values at the edge of the 40\degr\ by 40\degr\ images are 
about $1.4\times 10^{10}$ cm$^2$ s. Figure 3 \textit{(upper part)} shows the coadded counts, exposure and 
intensity map for the 50 cluster sample.\\

EDITOR place figure 3 here\\

We have analyzed these images using the maximum-likelihood technique as already described. Here, however, we have to 
provide a customized diffuse galactic gamma-ray foreground model, adapted for the application of our superpositioned 
cluster sample in the cluster--centered coordinate system. This was achieved by adopting the standard diffuse galactic 
emission model (GALDIF) \citep{hun97} used for EGRET likelihood analysis, given on a 0.5\degr\ by 0.5\degr\ grid, into a 
specific diffuse foreground model for the galaxy clusters in cluster--centered coordinates: Corresponding in image size and 
coordinates with the counts, exposure and intensity map of the individual clusters, the appropriate diffuse maps have been taken 
directly from the GALDIF model. These maps were subsequently transformed into the cluster--centered coordinate system. 

In each individual cluster observation we expect the diffuse foreground emission to contribute a certain number of counts, $c_{i}$,
that can be calculated as the product of the respective exposure map, $\varepsilon_{i}$, and the intensity of the diffuse emission, 
$DF_{i}$. The appropriate diffuse foreground intensity for the coadded data, $DF_{tot}$, is then derived by:

\begin{displaymath}
DF_{tot} = {\frac{1}{\varepsilon_{tot}}} {\sum_{i}{c_{i}}} = {\frac{1}{\varepsilon_{tot}}} {\sum_{i}{\varepsilon_{i}}DF_{i}}, \qquad \varepsilon_{tot} = {\sum_{i}{\varepsilon_{i}}}.
\end{displaymath}

By applying the maximum likelihood procedure in conjunction with the appropriately adapted diffuse model, 
the 40\degr\ by 40\degr\ images were searched for excesses. Of interest here is only the map center, 
corresponding to the emission maximum of the considered galaxy clusters in X-rays. No significant gamma-ray emission excess 
has been found within a radius of 5\degr\ of the origin in the cluster-centered image. With a cumulative 
exposure of 3.4 $\times$ $10^{10}$ cm$^{2}$ s for E $>$ 100 MeV the corresponding upper limit 
is 5.9 $\times$ $10^{-9}$ cm$^{-2}$ s$^{-1}$ (averaged over the four central 0.5\degr\ $\times$ 0.5\degr\ map bins) 
for the so--constructed average galaxy cluster. Fig.3 \textit{(lower part)} shows the customized diffuse model and the 
resulting likelihood test statistics maps. Easily seen in the test statistics image is 3C279, located at $\sim$ 13\degr\ from the 
map center, and thus far too distant to cause any conflict with the determined upper limit.

\section{Discussion}
\subsection{Cases of individual galaxy clusters}

The negative results from both an analysis of the gamma-ray data from the EGRET instrument at 
positions of 58 individual galaxy clusters as well as from a superposition of 50 galaxy 
clusters needs to be critically discussed with respect to underlying systematics. Categorically, 
the question of an appropriately chosen selection of galaxy clusters might arise. The assumption 
has been made that the brightest and nearest clusters detected at X-ray wavelengths should be the most 
likely candidates to emit observable gamma-rays, supported by various models explaining the 
multifrequency emission properties and the general understanding of confinement and interaction of 
cosmic rays in the intercluster medium (ICM) of a galaxy clusters \citep{ber97, voe96}. Because 
almost all clusters exhibiting unusual multifrequency emission characteristics (EUV-excess emission, 
non--thermal X-ray emission and/or a diffuse radio halo) are naturally included here, the above assumption 
is certainly not artificial. In Tab.2 we compare our results with two different scenarios of gamma-ray emission from 
cosmic ray interactions in the ICM, quantitatively predicting gamma-ray emission for some individual galaxy clusters.

EDITOR place table 2 here\\

Our upper limits are consistently below the predictions as given by \cite{ens97}, and especially \cite{dar95}. 
Thus, the suggested scenarios are ruled out in the given parameter space. Concerns against the results of 
\cite{dar95} have been already pointed out earlier by \cite{ste96b} on account of the spectra of the 
diffuse galactic and extragalactic gamma-ray background. With the apparent conflict between these model predictions 
and the observational upper limits for the individual galaxy clusters, 
we clearly disfavor these models against models which predict gamma-ray emission below the sensitivity of the 
instruments of the Compton gamma-ray observatory era like inverse Compton scenarios from cosmic ray electrons 
accelerated at accretion shocks by \citep{cola98, min02}.\\ 
At present, with quantitative predictions about the gamma-ray emission made only in cases of the most prominent 
galaxy clusters, and with measured upper limits and predicted fluxes not orders 
of magnitudes apart, only a moderate conservative interpretation is tenable:  
The degree of freedom in the parameters in the models in conflict with our measurements, especially
if predicting a diffuse extragalactic background component without a single positive detection of
the object class in question made does not allow to discriminate between the suggested scenarios solely 
on the basis of the presented gamma-ray data. 
Not until the next generation of gamma-ray telescope will be available with good prospects 
to detect individual galaxy clusters as well as the chance to observationally constraint the constituents of the 
extragalactic diffuse background, the sensitive measurements to discriminate the 
various scenarios for gamma-ray emission from galaxy clusters must originate at other wavelengths. 
Here, especially three wavebands are of particular interest: the new generation of low-threshold 
ground--based Cherenkov--telescopes (IACTs) presently being built or just starting their observations will 
provide very sensitive measurements at energies above 100 GeV, thus providing the necessary 
information to constrain the inverse Compton component at the highest energies; more and better 
hard X-ray measurements may clearly discriminate between thermal and non--thermal emission components 
and therefore help to identify the nature of the originating particle population, in particular deciding 
on the preference of inverse Compton or non--thermal Bremsstrahlung scenarios; finally, more sensitive 
radio halo measurements may be performed, especially at the high--frequency end of the synchrotron 
emission component, where currently only the radio halo of the Coma cluster has been sufficiently 
investigated \citep{sch87, thi02}. High-frequency radio measurements from diffuse cluster halos will 
directly constrain the shape and intensity of the resulting inverse Compton component at high energies. 


\subsection{The case of galaxy clusters as a population}

We now discuss the result for the average cluster from our analysis of 50 galaxy clusters population. 
At the achieved sensitivity there is still no indication of gamma-ray emission from galaxy clusters. 
The 2 $\sigma$ upper limit for the flux of the average cluster is 5.9 $\times$ $10^{-9}$ cm$^{-2}$ s$^{-1}$.
This may help to  resolve the stark inconsistency between studies performing direct EGRET data analysis on 
galaxy clusters (i.e. \cite{gly94}, \cite{sre96}, and this work) and the recently published detection claims 
originating from information on gamma-ray point sources from the 3EG catalog in conjunction with statistical 
assessments by \cite{cola01}, \cite{cola02} and \cite{kaw02}. Here, we would like to give a reassessment of 
these detection claims on a firm spatial--statistical basis. The cluster sample studied by Colafrancesco 
consists of the entire Abell cluster catalog \citep{abe89} at galactic latitudes $|b| > 20$\degr\ . 
The corresponding Poissonian probability distribution for spatial association between the cluster sample and gamma-ray 
point sources form the EGRET catalog is easy to determine and given in Fig.4a. 
The two-point autocorrelation function, $\omega_{\Theta}$, for the population of Abell clusters has been intensively 
studied previously \citep{hau73, pos86, oli90, aky00}, but its impact on the correlation analysis can be rigorously calculated
only for a very large sample of objects. Here we are concerned with only 170 unidentified EGRET sources, 59 of which are located
at $|b| > 20\degr$. Therefore, we have directly determined the chance probability for an association between an arbitrary source
and one of the Abell clusters as a function of the radius--of--interest, i.e. the maximum separation of sources considered as 
associated. \cite{cola02} reports an association of 50 EGRET sources (resp. 18 unidentified EGRET sources) and 70 Abell cluster 
(resp. 24 Abell clusters) based on a initial sample of 3979 Abell clusters and 128 EGRET sources (resp. 59 unidentified EGRET sources) 
at $|b| > 20\degr$. Pure Poissonian statistics predicts a total of 34.4 (17.2) single, 8.6 (2.0) double and 1.3 (0.2) triple 
associations between EGRET sources and Abell associations at 1\degr\ roi (resp. 0.81\degr\ as the average $\Theta_{95}$). 
Using the modified chance probabilities (Fig. 4b) to account for the autocorrelation of the Abell clusters, one expects a 
total of 28.0 (10.6) single, 8.9 (2.5) double and 2.6 (0.5) triple association by chance. Autocorrelation among the unidentified 
EGRET sources at $\leq$ 1\degr\ scale is per se excluded due to the moderate angular resolution of the EGRET telescope and its 
relatively inability to discriminate neighboring sources adequately ("source confusion"). Hence the expected chance associations 
amount to 40.7 EGRET sources and 56.6 Abell clusters for 128 EGRET sources at 1\degr\ roi and 13.7 EGRET sources and 17.5 Abell clusters 
for 59 unidentified EGRET sources at $\Theta_{95}$, respectively. In terms of the cumulative Poisson probability, the significance of 
the correlation claim by Colafrancesco needs to be reassessed to only 1.36 $\sigma$ in case of the 128 EGRET sources (which is anyway meaningless due to the contamination with 
already \textit{identified} EGRET sources, the blazars), or 1.03 $\sigma$ in case of the 59 unidentified EGRET sources, rigorously 
indicating the statistical insignificance of the correlation claim.   

EDITOR place figure 4 here\\

Furthermore, the explicitly suggested Abell cluster/unidentified EGRET source associations by \cite{cola02} do not contain 
the most likely and prominent galaxy clusters as predicted to be the clusters with the best chance to be detected in  
high-energy gamma rays (Fig.5). Actually, only one of the 18 listed unidentified EGRET sources indeed has a truly 
X-ray bright Abell cluster counterpart; the remaining 23 clusters are not even included in the most recent 
and carefully compiled sample of a flux-limited X-ray bright galaxy clusters population (HIFLUGCS: \cite{reip02}). 
Thus, the identification sequence as suggested by Colafrancesco (radio halo $\rightarrow$ X-ray brightness
$\rightarrow$ counterpart in unidentified gamma-ray source) simply does not work this way, neither globally nor 
just in a few of the referenced cases. Similarly, the predictive power of the L$_{\gamma}$--L$_{radio}$-- and 
L$_{\gamma}$--L$_{X}$--correlation as given in \cite{cola02} must be seriously questioned.\\

EDITOR place figure 5 here\\

A further suggestion of possibly merging clusters as counterparts of unidentified gamma-ray sources by 
\cite{kaw02} is based on only seven individual gamma-ray sources, all of which belong to the sample of steady 
unidentified EGRET sources classified as such by \cite{geh00}. However, more specific variability studies of EGRET 
sources as performed by \cite{tom99} and \cite{tor01} do not support non-variability for each of these sources.
Two sources clearly exhibit gamma-ray flux variability, and a further two sources belong to the group of gamma-ray  
sources where we could not decide on the basic of the current gamma-ray data whether or not a source is truly non--variable. 
Thus, a more conservative sample would consist of only three candidates, 
reducing the statistical significance of the remaining associations to the level of expected chance coincidences. 
Certainly, those newly-suggested counterparts are worthy of an intensive investigation in order to conclude about 
their true nature and, subsequently, about their chance to be related to an unidentified gamma-ray source. 
At present, this study does not leave sufficient confidence for an association of unidentified gamma-ray sources 
and cluster merger systems. Recently, the problem of such association has been theoretically addressed by \cite{berr02}, 
arguing that it is unlikely that more than a few of the isotropically distributed unidentified EGRET sources at high-galactic 
latitudes can be attributed to radiation from non--thermal particles produced by cluster merger shocks. Similar conclusions 
were reached by \cite{gab02}, emphasizing the discrepancy between non--thermal activity in galaxy clusters and the relative 
ineffectiveness of major shocks to energize the underlying particle population.\\

Yet another detection claim of galaxy clusters in the gamma-ray data was published recently. A population study 
by \cite{sch02} claims a statistical detection at $\geq 3\sigma$ level from unresolved gamma-ray sources 
spatially coincident with Abell clusters, in particular those of high optical richness. It quotes a mean cluster 
flux of $\sim$ 1.14 $\times$ $10^{-9}$ cm$^{-2}$ s$^{-1}$ at E $>$ 100 MeV in a 1\degr\ radius aperture, corresponding 
to a 68 \% flux enclosure of the EGRET point-spread function. However, the energy--averaged EGRET flux enclosure at 
E $>$ 100 MeV in a 1\degr\ aperture, assuming a source spectrum with a power-law index of -2.0, is in fact much lower, 
namely 24.1 \%. The 68 \% flux enclosure is reached at a radius of 3.1\degr\ . Therefore the flux quoted for the 
innermost (1$^{\circ}$) radial bin in the study, $1.19\times 10^{-6}$ ph s$^{-1}$ cm$^{-2}$ sr$^{-1}$, in fact corresponds 
to a total source flux of 4.7 $\times$ $10^{-9}$ cm$^{-2}$ s$^{-1}$, which is marginally inconsistent with our findings. 
Considering a 2\degr\ aperture by inclusion of the second innermost bin of $\sim$ 1.2 $\times 10^{-6}$ ph s$^{-1}$ cm$^{-2}$ sr$^{-1}$
from \cite{sch02}, the corresponding mean cluster flux will increase to 9.6 $\times$ $10^{-9}$ cm$^{-2}$ s$^{-1}$, 
which is clearly inconsistent with our upper limit for the average cluster from the sample of nearby, X-ray bright galaxy
clusters. Since it is not within the scope of this paper to perform an in--depth study of possible systematic effects, 
we only state that the results presented by \cite{sch02}, although at comparable sensitivity level, are apparently 
inconsistent with ours.\\     

In conclusion, we still have to await the first observational evidence for the high-energy gamma-ray emission
of galaxy clusters. The last generation of gamma-ray telescopes aboard CGRO was not able to resolve an individual 
galaxy cluster nor the nearby, X-ray brightest clusters of galaxies as a population. Until the next generation of 
gamma-ray instruments will challenge this important scientific topic, progress is expected at other wavelengths: 
from GHz-frequency radio observations of radio halos, from studies of soft and hard X-ray excess features with sufficient 
statistical significance, and from measurements of the new generation of imaging atmospheric Cherenkov telescopes. 
 
\acknowledgments

O.R. acknowledges support from a National Academy of Science/National Research Council 
Associateship at NASA/Goddard Space Flight Center, where part of this study was executed.
O.R. and M.P. are supported from DLR QV0002. O.R. would like to thank the many people he 
had consulted about the topic, in particular D.L. Bertsch, P. Blasi, S. Digel, F. Miniati, 
G. Madejski, V. Petrosian, D.J. Thompson, R. Schlickeiser and F. Stecker, for the general 
encouragement to continue this study beyond a conference contribution by reason of the 
rather unforeseen interest the topic experienced recently.

{}

\clearpage

\begin{figure}
\plotone{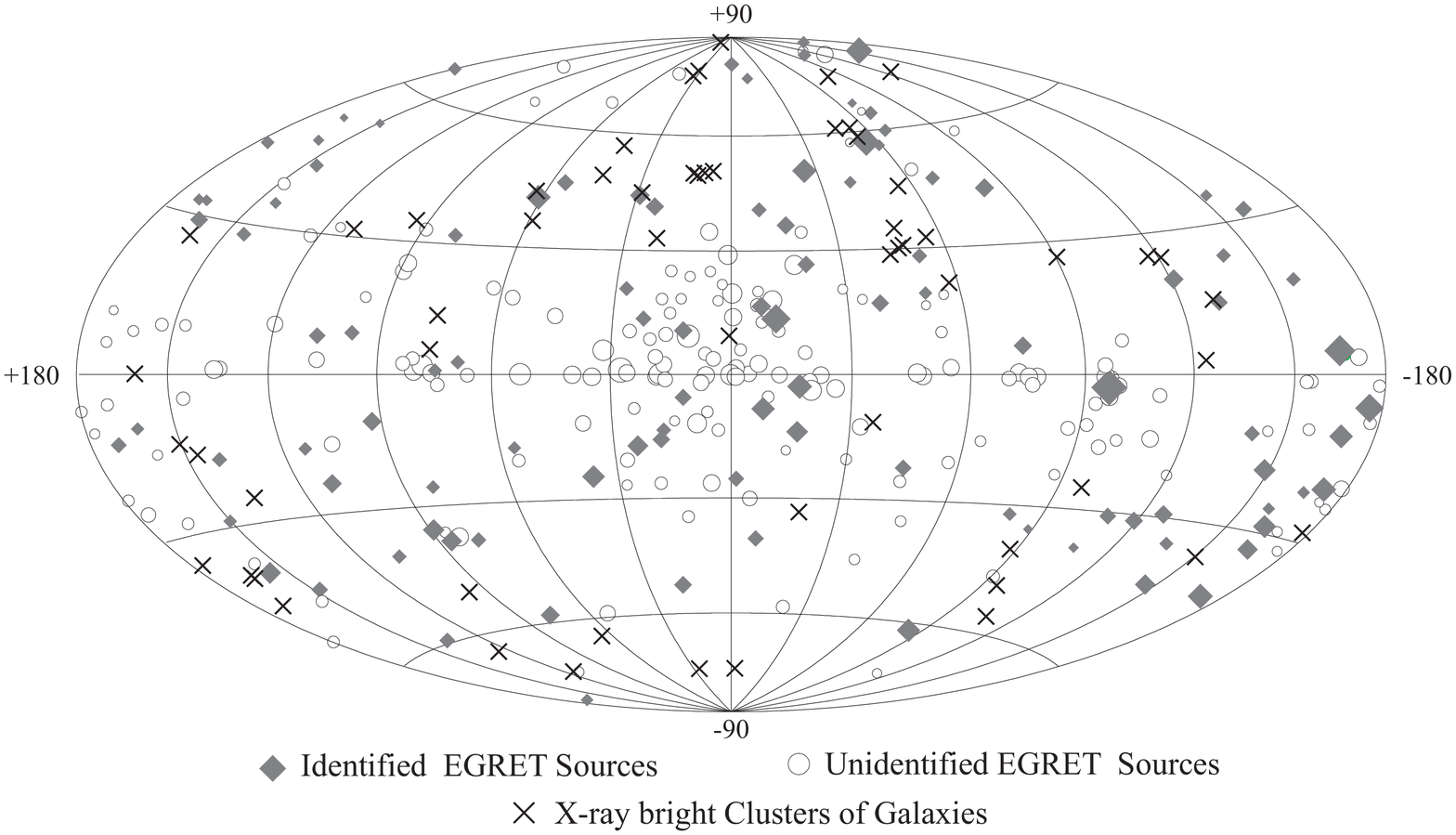}
\caption{Positions of 58 X-ray bright galaxy clusters ($z < 0.14$) and 
EGRET-detected gamma-ray sources ($E > 100$ MeV), in galactic coordinates.}
\label{fig1}
\end{figure} 

\clearpage

\begin{figure}
\plotone{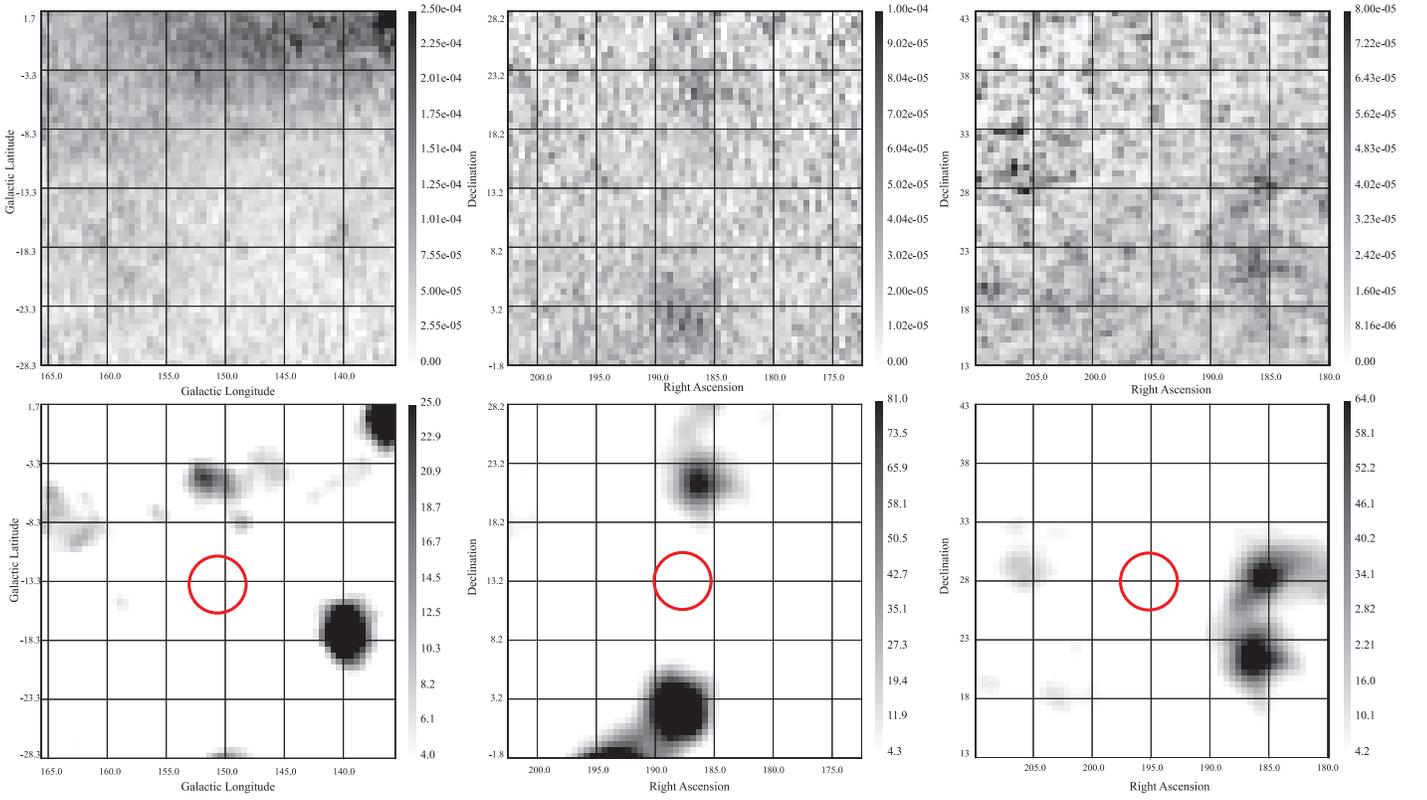}
\caption{Prominent individual galaxy clusters as observed by EGRET ($E > 100$ MeV). 
From left to right: The Perseus, Virgo, and Coma cluster, shown as gamma-ray intensity (in units of cm$^{-2}$ s$^{-1}$ sr$^{-1}$) 
and likelihood test statistics. The likelihood test statistics corresponds to the square of the source detection significance. 
The region of interest is indicated by the 2.5\degr\ circle, centered at the X-ray emission maximum of the galaxy cluster.}
\label{fig2}
\end{figure} 

\clearpage

\begin{figure}
\epsscale{0.75}
\plotone{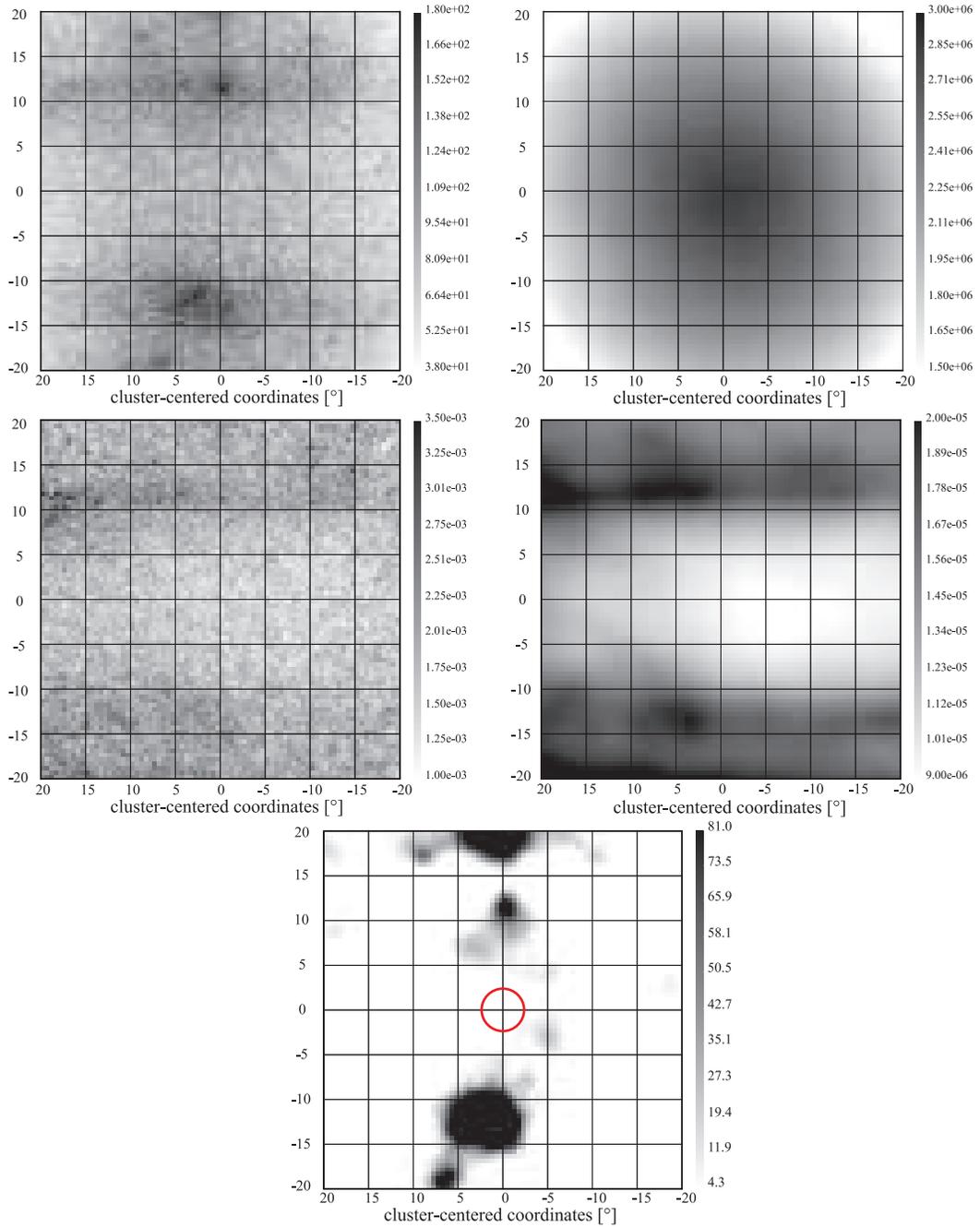}
\caption{Upper figures: Counts and exposure maps for the combined data set of 50 galaxy clusters as observed by EGRET ($E > 100$ MeV).
Counts are given in units of photons per pixel, exposure in units of cm$^{2}$ s. Center figures: Intensity map and diffuse
foreground model for the combined data set of 50 galaxy clusters as observed by EGRET ($E > 100$ MeV). Both are given in    
units of cm$^{-2}$ s$^{-1}$ sr$^{-1}$. The foreground model already includes the exposure weight factors of the individual 
galaxy clusters. Lower figure: Likelihood test statistics map (corresponding to the square of the source detection significance) 
for the combined data set of 50 galaxy clusters as observed by EGRET ($E > 100$ MeV). The region of interest is 
indicated by the 2.5\degr\ circle. All maps are shown in a cluster--centered coordinate system.}
\label{fig3}
\end{figure} 

\clearpage


\clearpage

\begin{figure}
\epsscale{1.0}
\plottwo{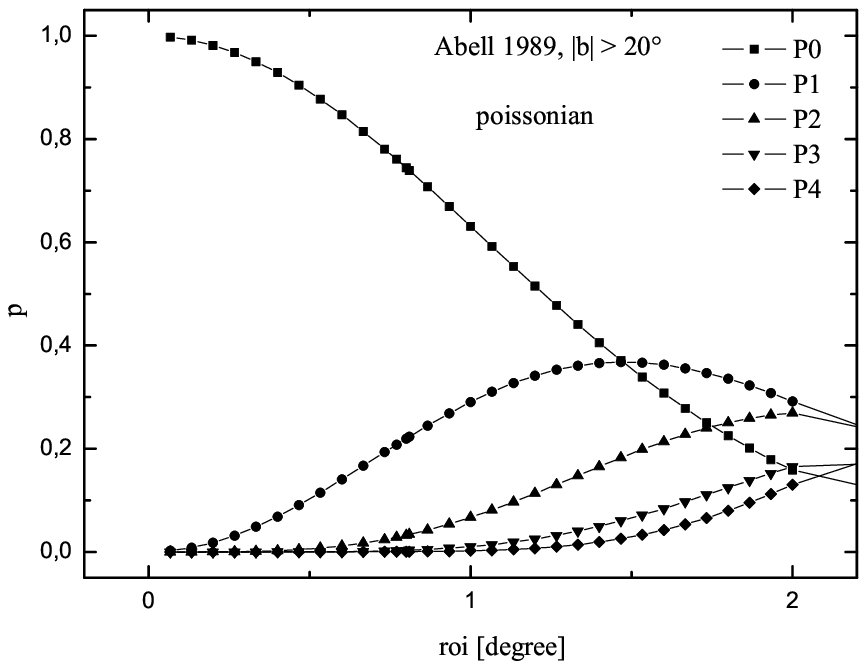}{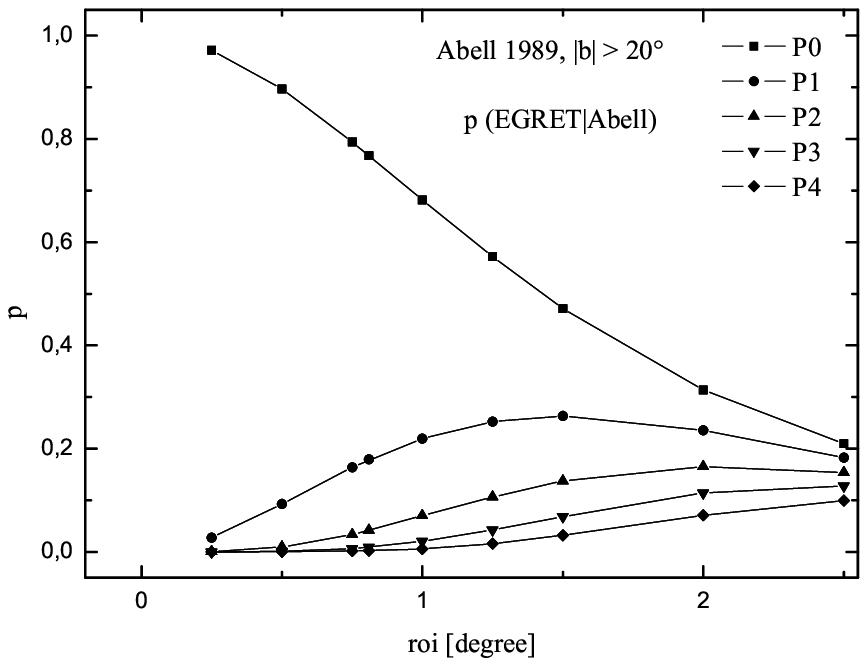}
\caption{Poissonian statistics and modified statistics for cluster--cluster--autocorrelation for 
Abell clusters at $|b| > 20\degr$. P0, P1, P2 etc. are the probabilities for none, single, double 
coincidences between an arbitrary source and the Abell cluster sample for different source separations.}
\label{fig4}
\end{figure} 

\clearpage

\begin{figure}
\epsscale{0.8}
\plotone{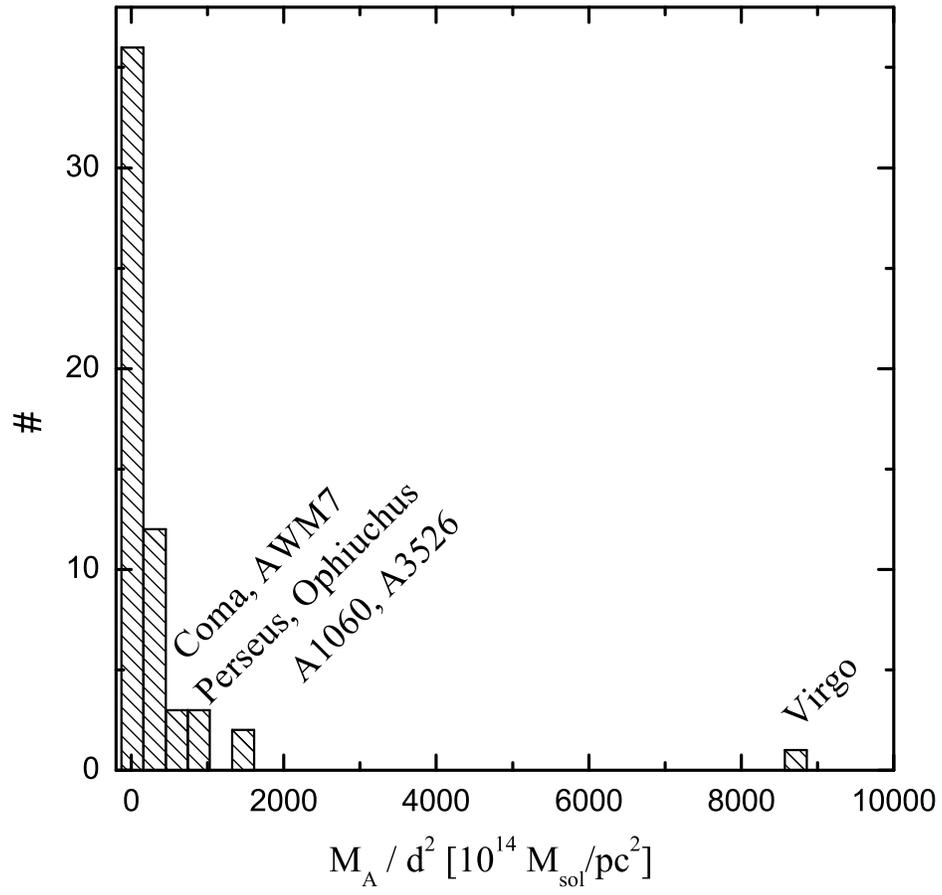}
\caption{Expected detectabiliy of galaxy clusters in gamma rays by means of a mass over distance$^{2}$ measure.
Cluster masses M$_{A}$ = M$_{tot}$($< r_{A}$) have been obtained from \cite{reip02}.}
\label{fig5}
\end{figure}

\clearpage

\begin{deluxetable}{lcrrrrll}
\tabletypesize{\tiny}
\tablecolumns{9}
\tablewidth{6.8in}
\tablenum{1}
\tablecaption{EGRET observations towards X-ray selected galaxy clusters}
\tablehead{
\colhead{$\#$} & \colhead{Name} & \colhead{l}            & \colhead{b}            & \colhead{z} & \colhead{flux ($>$100 MeV)}             & \colhead{viewing periods} & \colhead{remarks}  \\
\colhead{}     & \colhead{}     & \colhead{[$^{\circ}$]} & \colhead{[$^{\circ}$]} & \colhead{}  & \colhead{[$10^{-8}$ cm$^{-2}$~s$^{-1}$]} & \colhead{}                & \colhead{}
}

\startdata

1 & A426 (PER Cluster)	& 150.58 & -13.26 & 0.0184 & $<$ 3.72 & 0150,0310,0360,0365,0390,2110,3250,4270,7287,7289 & \\

2 & OPH Cluster			&   0.56 &   9.27 & 0.028  & $<$ 5.00 & 0050,0160,0270,2100,2140,2190,2230,2260,2290,2295,  & \\
  &                     &        &        &        &          & 2320,3023,3230,3240,3300,3320,3340,3365,4210,4220,  & \\
  &                     &        &        &        &          & 4230,4235,4290,5080,5295,6250,6151 & \\ 

3 & VIR Cluster			& 282.08 &	75.20 &	0.0038 & $<$ 2.18 & 0030,0040,0110,2040,2050,2060,3040,3050,3060,3070,  & \\
  &                     &        &        &        &          & 3080,3086,3110,3116,3120,3130,4050,4060,4070,4080,  & \\
  &                     &        &        &        &          & 5110,6105,6215,8065,8067,9100,9111 & \\

4 & COMA Cluster		&  58.13 &  88.01 & 0.0238 & $<$ 3.81 & 0030,0040,0110,2040,2050,2060,2180,2220,3040,3050,  & \\
  &                     &        &        &        &          & 3070,3080,3086,3110,3116,3120,3130,4060,4070,4180,  & \\
  &                     &        &        &        &          & 5150,7155 & \\

5 & A2319				&  75.68 &	13.50 & 0.056  & $<$ 3.79 & 0020,0071,2010,2020,2030,2120,3020,3032,3034,3037,  & \\
  &                     &        &        &        &          & 3181,3280,3310,3315,3330,7100,7110 & \\

6 & A3571				& 316.31 &	28.54 &	0.04   & $<$ 6.34 & 0120,0230,0320,2070,2080,2150,2170,3160,4050,4080, & \\						   
  &                     &        &        &        &          & 4240 & \\
  
7 & A3526 (CEN Cluster) & 302.40 &  21.55 &	0.0109 & $<$ 5.31 & 0120,0140,0230,0320,2070,2080,2150,2170,3030,3140,  & \\
  &                     &        &        &        &          & 3150,3160,4020,4025,4240 & \\

8 & TRA Cluster			& 324.36 & -11.38 &	0.051  & $<$ 8.13 & 0230,0270,0350,0380,2320,3140,3150,3365,4020,4025 & \\

9 & 3C129 (3A 0446+449) & 160.39 &   0.13 & 0.0223 & $<$ 5.29 & 0002,0005,0150,0310,0360,0365,0390,2130,2210,3211,  & \\
  &                     &        &        &        &          & 3215,3195,3250,4120,4260,4270 & \\

10 & AWM7 (2A 0251+413)	& 146.34 & -15.63 & 0.018  & $<$ 3.47 & 0150,0360,0365,0390,2110,3250,4270,7287,7289 & \\

11 & A754 	  			& 239.20 &  24.71 &	0.054  & $<$ 8.18 & 0300,0330,0410,0440 & \\

12 & A2029				&   6.49 &	50.55 &	0.0768 & $<$ 7.49 & 0240,0245,0250,3390,4060,4070 & \\

13 & A2142				&  44.23 &	48.69 &	0.0899 & $<$ 4.97 & 0092,0240,0245,0250,2010,2020,3034,3390,5165,5190,  & \\
   &                    &        &        &        &          & 7210,7225 & \\

14 & A2199				&  62.93 &	43.69 &	0.0299 & $<$ 9.27 & 0092,2010,2020,3034,4030,5156,5190,6178,7210,7225 & a \\

15 & A3667				& 340.88 & -33.39 & 0.055  & $<$ 3.82 & 0350,0380,0420,2090,3230 & \\

16 & A478				& 182.43 & -28.29 &	0.09   & $<$ 5.14 & 0002,0003,0004,0005,0010,0021,0210.0360,0365,0390, & \\
   &                    &        &        &        &          & 2130,2210,3211,3215,3170,3370,4120,4130,4191,4195, & \\
   &                    &        &        &        &          & 4200,4260 & \\

17 & A85				& 115.04 & -72.06 &	0.055  & $<$ 6.32 & 0091,0132,0210,3270,3360,4040,4250,4280,9150,9160 & b \\

18 & A3266				& 272.14 & -40.16 &	0.0545 & $<$ 4.42 & 0060,0100,0170,2200,2240,3290,3350,3355,4090,4150, & \\
   &                    &        &        &        &          & 5210 & \\

19 & A401				& 164.18 & -38.87 & 0.075  & $<$ 9.28 & 0150,0210,0360,0365,0390,3170,4250,6311,9175 & \\

20 & 3A 0745-191		& 236.42 &   2.99 &	0.1028 & $<$ 7.08 & 0007,0080,0410,0440,3010,3385,5100,5105 & \\

21 & A496				& 209.57 & -36.48 &	0.0327 & $<$ 7.11 & 0290,3370,4191,4195,4200 & \\

22 & A1795				&  33.81 &  77.18 &	0.063  & $<$ 3.98 & 0030,0040,0240,0245,2180,2220,3070,3080,3086,3110, & \\
   &                    &        &        &        &          & 3116,3120,3130,4060,4070,5150 & c \\

23 & A2256				& 111.10 &	31.74 &	0.056  & $<$ 4.28 & 0180,0220,2160,3190,4010 & \\

24 & CYG A Cluster		&  76.19 &	 5.76 &	0.057  & $<$ 4.46 & 0020,0071,2030,2120,3020,3032,3034,3037,3181,3280, & \\
   &                    &        &        &        &          & 3310,3315,3330,6011 & \\

25 & 2A 0335+096		& 176.25 & -35.08 &	0.0349 & $<$ 8.11 & 0210,0360,0365,0390,2130,3170,4200,6311,9175 & \\

26 & A1060				& 269.63 &  26.50 &	0.0114 & $<$ 14.85 &	0007,0300,0320,0330,2300,2305,3010,3030,3385 & \\

27 & A3558				& 312.00 &	30.72 &	0.048  & $<$ 3.58 & 0120,0230,0320,2070,2080,2150,2170,3160,4050,4080, & \\
   &                    &        &        &        &          & 4240 & \\

28 & A644				& 229.93 &	15.29 &	0.0704 & $<$ 9.71 & 0300,0330,0410,0440,4035,5100,5105 & d  \\

29 & A1651				& 306.73 &	58.63 &	0.086  & $<$ 3.75 & 0030,0110,2040,2050,2060,2070,3040,3050,3060,3070, & \\
   &                    &        &        &        &          & 3080,3086,3110,3116,3120,3130,4050,4060,4070,4080, & \\
   &                    &        &        &        &          & 5110,5115,6060,6070,6080,6090,6100,6105,6111,6215, & \\
   &                    &        &        &        &          & 8065,8067,9100,9111 & e \\

30 & A3562				& 313.30 &	30.35 &	0.0499 & $<$ 3.62 & 0120,0230,0320,2070,2080,2150,2170,3160,4050,4080, & \\
   &                    &        &        &        &          & 4240 & \\

31 & A1367				& 234.80 &	73.03 &	0.0215 & $<$ 2.72 & 0030,0040,0110,2040,2050,2060,2180,2220,3040,3050, & \\
   &                    &        &        &        &          & 3060,3070,3080,3086,3110,3116,3120,3130,3220,3260, & \\
   &                    &        &        &        &          & 4180,5150,7155 & \\

32 & A399				& 164.36 & -39.46 &	0.072  & $<$ 4.92 & 0150,0210,0360,3170,4250,6311,9175 & \\

33 & A2147				&  28.80 &  44.49 &	0.0356 & $<$ 7.45 & 0092,0240,0245,0250,2010,3390 & f \\

34 & A119				& 125.74 & -64.11 &	0.044  & $<$ 4.51 & 0210,0260,0280,0370,3170,3200,3270,3360,4250 & \\

35 & A3158				& 264.68 & -48.76 &	0.0575 & $<$ 2.52 & 0060,0100,0170,0290,2200,2240,3290,3350,3355,4090, & \\
   &                    &        &        &        &          & 4150,5170,5210,8010,8020,8339 & \\

36 & HYD A Cluster		& 242.93 &  25.09 & 0.0538 & $<$ 7.24 & 0300,0330,0410,0440 & \\

37 & A2065 				&  42.86 &  56.56 &	0.06   & $<$ 5.51 & 0092,0240,0245,0250,2010,2020,3390 & \\

38 & A2052				&   9.42 &	50.12 &	0.0348 & $<$ 6.24 & 0240,0245,0250,3390,4060,4070 & \\

39 & A2063				&  12.82 &	49.69 &	0.036  & $<$ 5.52 & 0240,0245,0250,3390,4060,4070 & \\

40 & A1644				& 304.89 &	45.44 & 0.0456 & $<$ 2.89 & 0030,0110,0120,0320,2040,2050,2060,2070,2080,2150, & \\
   &                    &        &        &        &          & 2170,3040,3050,3060,3070,3120,3160,4050,4060,4070, & \\
   &                    &        &        &        &          & 4080,4240,5110,5115,6060,6070,6080,6090,6100,6105, & \\
   &                    &        &        &        &          & 6111,6215,8065,8067,9100 & \\

41 & Klem 44 (A4038)	&  25.08 & -75.90 &	0.0283 & $<$ 3.60 & 0091,0132,4040,4280 & \\

42 & A262 	 			& 136.58 & -25.09 &	0.0161 & $<$ 6.00 & 0150,0260,0280,2110,3170,3250,4250,4270,7287,7289 & \\

43 & A2204				&  21.09 &  33.25 &	0.153  & $<$ 7.99 & 0160,0240,0245,0250,3240,3390,4290 & \\

44 & A2597				&  65.36 & -64.84 &	0.0824 & $<$ 8.19 & 0091,0132,0190,3200,3220,3270,4040 & \\

45 & A1650				& 306.72 &  61.06 &	0.084  & $<$ 3.07 & 0030,0110,2040,2050,2060,2070,3040,3050,3060,3070, & \\
   &                    &        &        &        &          & 3080,3086,3110,3116,4050,4060,4070,4080,5110,5115, & \\
   &                    &        &        &        &          & 6060,6070,6080,6090,6100,6105,6111,6215,8065,8067, & \\
   &                    &        &        &        &          & 9100,9111 & \\

46 & A3112				& 252.95 & -56.09 &	0.0746 & $<$ 4.86 & 0100,0290,3290,3350,3355,4090,4280,5170,8010,8020, & \\
   &                    &        &        &        &          & 8339 & \\

47 & A3532				& 304.44 &  32.48 &	0.0537 & $<$ 7.42 & 0120,0320,2050,2070,2080,2150,2170,3160,4050,4080, & \\
   &                    &        &        &        &          & 4240 & \\
 
48 & A4059				& 356.84 & -76.06 &	0.0748 & $<$ 2.86 & 0091,0132,4040,4280 & \\

49 & A3395				& 263.18 & -25.13 &	0.0498 & $<$ 3.28 & 0007,0060,0080,0170,2300,3010,3290,3350,3355,3385, & \\
   &                    &        &        &        &          & 4090,4150,5210 & \\

50 & MKW 3s				&  11.38 &  49.45 &	0.0434 & $<$ 5.31 & 0240,0245,0250,3390,4060,4070 & \\

51 & A1689				& 313.38 &	61.10 &	0.1832 & $<$ 4.00 & 0030,0110,0240,0245,2040,2050,2060,2070,3040,3050, & \\
   &                    &        &        &        &          & 3060,3070,3080,5110,5115,6060,6070,6080,6090,6100, & \\
   &                    &        &        &        &          & 6105,6111,6215,8065,8067,9100,9111 & \\

52 & A576				& 161.42 &	26.24 &	0.038  & $<$ 3.47 & 0006,0180,0310,2160,2270,2280,3190,3195,4111,4115, & \\
   &                    &        &        &        &          & 5185 & \\

53 & A2244				&  58.81 &	36.31 &	0.097  & $<$ 4.29 & 0092,2010,2020,3034,4030,5165,5190,6178,7210,7225 & \\ 

54 & A2255				&  93.95 &	34.93 &	0.0808 & $<$ 5.43 & 0092,0220,2010,2020,2120,3020,3032,3034,3037,4030, & \\
   &                    &        &        &        &          & 7100,7110 & \\

55 & A1736				& 312.55 &	35.10 &	0.0431 & $<$ 3.48 & 0120,0320,2040,2050,2060,2070,2080,2150,2170,3160, & \\
   &                    &        &        &        &          & 4050,4080,4240,5115 & \\

56 & A400				& 170.24 & -44.94 &	0.0238 & $<$ 6.51 & 0210,3170,4250,6311,9175 & \\

57 & A2657				&  96.65 & -50.30 &	0.0414 & $<$ 7.43 & 0190,0260,0280,0370,3200,3270,3360,4100,4250,5070, & \\
   &                    &        &        &        &          & 5075 & \\

58 & A1775				&  31.99 &  78.73 &	0.0722 & $<$ 3.47 & 0030,0040,0110,0240,0245,2180,2220,3040,3070,3080, & \\
   &                    &        &        &        &          & 3086,3110,3116,3120,3130,4060,4070,5150 & \\

\enddata

\tablecomments{
a   $\sim 1.9 \degr$ of 3EG J1635+3813  \\
b	$\sim 0.8 \degr$ of 3EG J0038-0949  \\
c	$\sim 3 \degr$ of 3EG J1347+2932  \\
d	$\sim 1.4 \degr$ of 3EG J0812-0646 \\
e	$\sim 1.8 \degr$ of 3EG J1255-0549 \\
f	$\sim 0.7 \degr$ of 3EG J1605+1553 \\
}

\end{deluxetable}

\clearpage

\begin{deluxetable}{lccc}
\tabletypesize{\small}
\tablewidth{7.0in}
\tablecolumns{4}
\tablenum{2}
\tablecaption{\label{}Comparison of the EGRET upper limits with model predicted gamma-ray fluxes}\

\tablehead{
\colhead{Cluster} &
\colhead{$F_{\gamma}$ this measuremnt } &
\colhead{$F_{\gamma}$ from En{\ss}lin et al.(1997) }  &
\colhead{$F_{\gamma}$ from Dar \& Shaviv (1995) }
}

\startdata
 A426 (Perseus)   &$ < 3.7 \times 10^{-8}$ &$12 \times 10^{-8}$& $10 \times 10^{-8}$ \\
 Ophiuchus        &$ < 5 \times 10^{-8}$ &$9 \times 10^{-8}$ &  \nodata             \\
 A1656 (Coma)     &$ < 3.8\times 10^{-8}$ &$6 \times 10^{-8}$&$5 \times 10^{-8}$ \\
 M87 (Virgo)      &$ < 2.2\times 10^{-8}$ &$3 \times 10^{-8}$& $22 \times 10^{-8}$

\enddata
\tablecomments{$F_{\gamma}$ in units of ph ${\rm cm^{-1}\,s^{-1}}$}

\end{deluxetable}

\end{document}